\documentclass{article}
\pdfoutput=1 
\usepackage{url,subcaption}
\usepackage[onehalfspacing]{setspace}
\usepackage{dirtree}

\usepackage{fullpage}
\usepackage[T1]{fontenc}
\usepackage{ae}
\usepackage{aecompl}
\usepackage{pifont,array}

\usepackage[multiple]{footmisc}

\usepackage{authblk}
\usepackage{fancyvrb}
\usepackage{color}

\makeatletter
\def\PY@reset{\let\PY@it=\relax \let\PY@bf=\relax%
    \let\PY@ul=\relax \let\PY@tc=\relax%
    \let\PY@bc=\relax \let\PY@ff=\relax}
\def\PY@tok#1{\csname PY@tok@#1\endcsname}
\def\PY@toks#1+{\ifx\relax#1\empty\else%
    \PY@tok{#1}\expandafter\PY@toks\fi}
\def\PY@do#1{\PY@bc{\PY@tc{\PY@ul{%
    \PY@it{\PY@bf{\PY@ff{#1}}}}}}}
\def\PY#1#2{\PY@reset\PY@toks#1+\relax+\PY@do{#2}}

\expandafter\def\csname PY@tok@gd\endcsname{\def\PY@tc##1{\textcolor[rgb]{0.63,0.00,0.00}{##1}}}
\expandafter\def\csname PY@tok@gu\endcsname{\let\PY@bf=\textbf\def\PY@tc##1{\textcolor[rgb]{0.50,0.00,0.50}{##1}}}
\expandafter\def\csname PY@tok@gt\endcsname{\def\PY@tc##1{\textcolor[rgb]{0.00,0.27,0.87}{##1}}}
\expandafter\def\csname PY@tok@gs\endcsname{\let\PY@bf=\textbf}
\expandafter\def\csname PY@tok@gr\endcsname{\def\PY@tc##1{\textcolor[rgb]{1.00,0.00,0.00}{##1}}}
\expandafter\def\csname PY@tok@cm\endcsname{\let\PY@it=\textit\def\PY@tc##1{\textcolor[rgb]{0.25,0.50,0.50}{##1}}}
\expandafter\def\csname PY@tok@vg\endcsname{\def\PY@tc##1{\textcolor[rgb]{0.10,0.09,0.49}{##1}}}
\expandafter\def\csname PY@tok@vi\endcsname{\def\PY@tc##1{\textcolor[rgb]{0.10,0.09,0.49}{##1}}}
\expandafter\def\csname PY@tok@vm\endcsname{\def\PY@tc##1{\textcolor[rgb]{0.10,0.09,0.49}{##1}}}
\expandafter\def\csname PY@tok@mh\endcsname{\def\PY@tc##1{\textcolor[rgb]{0.40,0.40,0.40}{##1}}}
\expandafter\def\csname PY@tok@cs\endcsname{\let\PY@it=\textit\def\PY@tc##1{\textcolor[rgb]{0.25,0.50,0.50}{##1}}}
\expandafter\def\csname PY@tok@ge\endcsname{\let\PY@it=\textit}
\expandafter\def\csname PY@tok@vc\endcsname{\def\PY@tc##1{\textcolor[rgb]{0.10,0.09,0.49}{##1}}}
\expandafter\def\csname PY@tok@il\endcsname{\def\PY@tc##1{\textcolor[rgb]{0.40,0.40,0.40}{##1}}}
\expandafter\def\csname PY@tok@go\endcsname{\def\PY@tc##1{\textcolor[rgb]{0.53,0.53,0.53}{##1}}}
\expandafter\def\csname PY@tok@cp\endcsname{\def\PY@tc##1{\textcolor[rgb]{0.74,0.48,0.00}{##1}}}
\expandafter\def\csname PY@tok@gi\endcsname{\def\PY@tc##1{\textcolor[rgb]{0.00,0.63,0.00}{##1}}}
\expandafter\def\csname PY@tok@gh\endcsname{\let\PY@bf=\textbf\def\PY@tc##1{\textcolor[rgb]{0.00,0.00,0.50}{##1}}}
\expandafter\def\csname PY@tok@ni\endcsname{\let\PY@bf=\textbf\def\PY@tc##1{\textcolor[rgb]{0.60,0.60,0.60}{##1}}}
\expandafter\def\csname PY@tok@nl\endcsname{\def\PY@tc##1{\textcolor[rgb]{0.63,0.63,0.00}{##1}}}
\expandafter\def\csname PY@tok@nn\endcsname{\let\PY@bf=\textbf\def\PY@tc##1{\textcolor[rgb]{0.00,0.00,1.00}{##1}}}
\expandafter\def\csname PY@tok@no\endcsname{\def\PY@tc##1{\textcolor[rgb]{0.53,0.00,0.00}{##1}}}
\expandafter\def\csname PY@tok@na\endcsname{\def\PY@tc##1{\textcolor[rgb]{0.49,0.56,0.16}{##1}}}
\expandafter\def\csname PY@tok@nb\endcsname{\def\PY@tc##1{\textcolor[rgb]{0.00,0.50,0.00}{##1}}}
\expandafter\def\csname PY@tok@nc\endcsname{\let\PY@bf=\textbf\def\PY@tc##1{\textcolor[rgb]{0.00,0.00,1.00}{##1}}}
\expandafter\def\csname PY@tok@nd\endcsname{\def\PY@tc##1{\textcolor[rgb]{0.67,0.13,1.00}{##1}}}
\expandafter\def\csname PY@tok@ne\endcsname{\let\PY@bf=\textbf\def\PY@tc##1{\textcolor[rgb]{0.82,0.25,0.23}{##1}}}
\expandafter\def\csname PY@tok@nf\endcsname{\def\PY@tc##1{\textcolor[rgb]{0.00,0.00,1.00}{##1}}}
\expandafter\def\csname PY@tok@si\endcsname{\let\PY@bf=\textbf\def\PY@tc##1{\textcolor[rgb]{0.73,0.40,0.53}{##1}}}
\expandafter\def\csname PY@tok@s2\endcsname{\def\PY@tc##1{\textcolor[rgb]{0.73,0.13,0.13}{##1}}}
\expandafter\def\csname PY@tok@nt\endcsname{\let\PY@bf=\textbf\def\PY@tc##1{\textcolor[rgb]{0.00,0.50,0.00}{##1}}}
\expandafter\def\csname PY@tok@nv\endcsname{\def\PY@tc##1{\textcolor[rgb]{0.10,0.09,0.49}{##1}}}
\expandafter\def\csname PY@tok@s1\endcsname{\def\PY@tc##1{\textcolor[rgb]{0.73,0.13,0.13}{##1}}}
\expandafter\def\csname PY@tok@dl\endcsname{\def\PY@tc##1{\textcolor[rgb]{0.73,0.13,0.13}{##1}}}
\expandafter\def\csname PY@tok@ch\endcsname{\let\PY@it=\textit\def\PY@tc##1{\textcolor[rgb]{0.25,0.50,0.50}{##1}}}
\expandafter\def\csname PY@tok@m\endcsname{\def\PY@tc##1{\textcolor[rgb]{0.40,0.40,0.40}{##1}}}
\expandafter\def\csname PY@tok@gp\endcsname{\let\PY@bf=\textbf\def\PY@tc##1{\textcolor[rgb]{0.00,0.00,0.50}{##1}}}
\expandafter\def\csname PY@tok@sh\endcsname{\def\PY@tc##1{\textcolor[rgb]{0.73,0.13,0.13}{##1}}}
\expandafter\def\csname PY@tok@ow\endcsname{\let\PY@bf=\textbf\def\PY@tc##1{\textcolor[rgb]{0.67,0.13,1.00}{##1}}}
\expandafter\def\csname PY@tok@sx\endcsname{\def\PY@tc##1{\textcolor[rgb]{0.00,0.50,0.00}{##1}}}
\expandafter\def\csname PY@tok@bp\endcsname{\def\PY@tc##1{\textcolor[rgb]{0.00,0.50,0.00}{##1}}}
\expandafter\def\csname PY@tok@c1\endcsname{\let\PY@it=\textit\def\PY@tc##1{\textcolor[rgb]{0.25,0.50,0.50}{##1}}}
\expandafter\def\csname PY@tok@fm\endcsname{\def\PY@tc##1{\textcolor[rgb]{0.00,0.00,1.00}{##1}}}
\expandafter\def\csname PY@tok@o\endcsname{\def\PY@tc##1{\textcolor[rgb]{0.40,0.40,0.40}{##1}}}
\expandafter\def\csname PY@tok@kc\endcsname{\let\PY@bf=\textbf\def\PY@tc##1{\textcolor[rgb]{0.00,0.50,0.00}{##1}}}
\expandafter\def\csname PY@tok@c\endcsname{\let\PY@it=\textit\def\PY@tc##1{\textcolor[rgb]{0.25,0.50,0.50}{##1}}}
\expandafter\def\csname PY@tok@mf\endcsname{\def\PY@tc##1{\textcolor[rgb]{0.40,0.40,0.40}{##1}}}
\expandafter\def\csname PY@tok@err\endcsname{\def\PY@bc##1{\setlength{\fboxsep}{0pt}\fcolorbox[rgb]{1.00,0.00,0.00}{1,1,1}{\strut ##1}}}
\expandafter\def\csname PY@tok@mb\endcsname{\def\PY@tc##1{\textcolor[rgb]{0.40,0.40,0.40}{##1}}}
\expandafter\def\csname PY@tok@ss\endcsname{\def\PY@tc##1{\textcolor[rgb]{0.10,0.09,0.49}{##1}}}
\expandafter\def\csname PY@tok@sr\endcsname{\def\PY@tc##1{\textcolor[rgb]{0.73,0.40,0.53}{##1}}}
\expandafter\def\csname PY@tok@mo\endcsname{\def\PY@tc##1{\textcolor[rgb]{0.40,0.40,0.40}{##1}}}
\expandafter\def\csname PY@tok@kd\endcsname{\let\PY@bf=\textbf\def\PY@tc##1{\textcolor[rgb]{0.00,0.50,0.00}{##1}}}
\expandafter\def\csname PY@tok@mi\endcsname{\def\PY@tc##1{\textcolor[rgb]{0.40,0.40,0.40}{##1}}}
\expandafter\def\csname PY@tok@kn\endcsname{\let\PY@bf=\textbf\def\PY@tc##1{\textcolor[rgb]{0.00,0.50,0.00}{##1}}}
\expandafter\def\csname PY@tok@cpf\endcsname{\let\PY@it=\textit\def\PY@tc##1{\textcolor[rgb]{0.25,0.50,0.50}{##1}}}
\expandafter\def\csname PY@tok@kr\endcsname{\let\PY@bf=\textbf\def\PY@tc##1{\textcolor[rgb]{0.00,0.50,0.00}{##1}}}
\expandafter\def\csname PY@tok@s\endcsname{\def\PY@tc##1{\textcolor[rgb]{0.73,0.13,0.13}{##1}}}
\expandafter\def\csname PY@tok@kp\endcsname{\def\PY@tc##1{\textcolor[rgb]{0.00,0.50,0.00}{##1}}}
\expandafter\def\csname PY@tok@w\endcsname{\def\PY@tc##1{\textcolor[rgb]{0.73,0.73,0.73}{##1}}}
\expandafter\def\csname PY@tok@kt\endcsname{\def\PY@tc##1{\textcolor[rgb]{0.69,0.00,0.25}{##1}}}
\expandafter\def\csname PY@tok@sc\endcsname{\def\PY@tc##1{\textcolor[rgb]{0.73,0.13,0.13}{##1}}}
\expandafter\def\csname PY@tok@sb\endcsname{\def\PY@tc##1{\textcolor[rgb]{0.73,0.13,0.13}{##1}}}
\expandafter\def\csname PY@tok@sa\endcsname{\def\PY@tc##1{\textcolor[rgb]{0.73,0.13,0.13}{##1}}}
\expandafter\def\csname PY@tok@k\endcsname{\let\PY@bf=\textbf\def\PY@tc##1{\textcolor[rgb]{0.00,0.50,0.00}{##1}}}
\expandafter\def\csname PY@tok@se\endcsname{\let\PY@bf=\textbf\def\PY@tc##1{\textcolor[rgb]{0.73,0.40,0.13}{##1}}}
\expandafter\def\csname PY@tok@sd\endcsname{\let\PY@it=\textit\def\PY@tc##1{\textcolor[rgb]{0.73,0.13,0.13}{##1}}}

\def\PYZus{\char`\_}

\def\PYZsq{\char`\'}
\def\PYZdq{\char`\"}

\makeatother

\author[1]{K. Jarrod Millman}
\author[2]{Matthew Brett}
\author[3]{Ross Barnowski}
\author[4]{Jean-Baptiste Poline}
\affil[1]{Division of Biostatistics, UC Berkeley, Berkeley, CA, USA}
\affil[1]{Berkeley Institute for Data Science, UC Berkeley, Berkeley, CA, USA}
\affil[2]{College of Life and Environmental Sciences, University of Birmingham, Birmingham, United Kingdom}
\affil[3]{Applied Nuclear Physics Program, Lawrence Berkeley National Laboratory, Berkeley, CA, USA}
\affil[4]{Neurology and Neurosurgery, McGill University, Montreal, Quebec, Canada}

\newcommand{\blockpar}[1]{\vspace*{3mm} \noindent \textbf{#1}}

\usepackage{amssymb}

\begin{document}
\onecolumn

\title{Teaching computational reproducibility for neuroimaging}

\maketitle

\begin{abstract} %
We describe a project-based introduction to reproducible and collaborative
neuroimaging analysis.
Traditional teaching on neuroimaging usually consists of a series of lectures
that emphasize the big picture rather than the foundations on which
the techniques are based. The lectures are often paired with practical
workshops in which
students run imaging analyses using the graphical interface of specific
neuroimaging software packages.   Our experience suggests that this
combination leaves the student with a superficial understanding of the
underlying ideas, and an informal, inefficient, and inaccurate approach to
analysis.  To address these problems, we based our course around a substantial
open-ended group project.  This allowed us to teach:
(a) computational tools to ensure
computationally reproducible work, such as the Unix command line, structured
code, version control, automated testing, and code review and
(b) a clear understanding of the statistical techniques used for a basic
analysis of a single run in an MRI scanner.
The emphasis we put on the group project showed the importance of
standard computational tools for accuracy, efficiency,
and collaboration.  The projects were broadly successful in engaging students
in working reproducibly on real scientific questions.  We propose that a
course on this model should be the foundation for future programs in
neuroimaging.  We believe it will also serve as a model for teaching efficient
and reproducible research in other fields of computational science.\\

\end{abstract}

\section{Introduction}

Few neuroimaging analyses are computationally reproducible,\footnote{Following
\cite{buckheit1995wavelab}, we define an analysis as \emph{computationally
reproducible} when someone other than the original author of an analysis can
produce on their own computer the reported results using the author's data,
code, and instructions.}
even between researchers in a single lab.
The process of analysis is typically ad-hoc and informal; the final result
is often the fruit of considerable trial and error, based
on scripts and data structures that the author inherited from other members of
the lab.
This process is
(a) confusing, leading to unclear hypotheses and conclusions,
(b) error prone, leading to incorrect conclusions and greater confusion,
and (c) an impractical foundation on which to build reproducible analyses.

Confusion, error, and lack of reproducibility are ubiquitous problems that
programmers have been fighting since before ``programmer'' became a job title.
There are widely accepted tools and processes to reduce these problems,
including the Unix command line, version control, high-level readable
programming languages, structured code, code review, writing tests for new
code, and continuous automatic test execution.

Many researchers accept that learning these techniques is desirable, but
believe that teaching them is too difficult, or would cost too much in class
time that should be spent on topics specific to neuroimaging.

In the course we describe here, we tested our hypothesis that we could
effectively teach \emph{both} the tools for efficient and reproducible work
\emph{and} the principles of neuroimaging, by building the course around a
substantial collaborative project, and putting the tools into immediate
practice.

At intake, our students had little or no prior exposure to neuroimaging,
or to the computational tools and process we listed above.  We set them the
open-ended task of designing and executing a project, which was either a
replication or an extension of a published neuroimaging analysis, built from
code they had written themselves.  We required the analysis to be computationally
reproducible; each project had to provide a short text file that gave a
simple set of commands with which the grading instructor could fetch the data,
run the analysis, and generate the final report, including figures.

\section{Background and Rationale}\label{background}

Between us, we have many decades experience of teaching neuroimaging analysis.
Like most other teachers of imaging, we have taught traditional courses with a
mixture of lectures covering the general ideas of the analysis, combined with
practical workshops using one of the standard imaging software packages.

We also have many years of experience giving practical support for imaging
analysis to graduate students and post docs.

Over these years of teaching and support, we have come to realize
that the traditional form of teaching does a poor job of preparing
students for a productive career in imaging research.  It fails in two
fundamental ways:

\begin{itemize}

\item
    Computation with large complex datasets is difficult and distracting;
        without training in standard practice to reduce this distraction, we
        condemn our students to a life-time of inefficient work and slow
        learning.

\item
    Standard teaching assumes either that the students already understand
        signal processing and linear algebra, or that they do not need to
        understand them.  In our experience, both of these assumptions are
        mostly false, with the result that the students do not have the
        foundation on which they can build further understanding.

\end{itemize}

As a result, imaging researchers usually do not have the vocabulary,
understanding, or shared tools to collaborate fluently with researchers from
other fields, such as statistics, engineering, or computer science.

The essential error in traditional teaching is one of emphasis;
it gives priority to the overview at the expense of the
intellectual and practical foundations.  The assumption is
that the student will either know or pick up the mathematical and
practical basis, with the big picture as a framework to guide them in
choosing what to learn.

In fact we find that it is rare for students to go on to learn these
foundations. They usually continue as they were when they leave the course:
as inefficient practitioners with little ability to reason about their
analysis.

In contrast to the traditional approach, we emphasize:

\begin{itemize}

\item
    Efficient computational practice to decrease confusion, reduce error, and
    facilitate collaboration.

\item
    The fundamental mathematical and statistical bases of the analysis, such
    as the linear model, using computational tools to build, illustrate, and
    explain.

\end{itemize}

We were inspired by the famous epithet of Richard Feynman, found written on
his blackboard after his death: ``What I cannot create, I do not
understand.''\footnote{\url{http://archives.caltech.edu/pictures/1.10-29.jpg}}
We first taught the students to build code efficiently, and then we taught
them how the methods worked, by building them from code.  Our aim was that our
students should be able to implement their own simple versions of the major
algorithms they were using.

Our course put great emphasis on a final group project that accounted for more
than half of the overall course grade.  We did this for two reasons:

First, we believe the motivations for computational reproducibility and
the difficulties of collaboration are too abstract to be meaningfully
understood in the absence of a significant, concrete, group project.
If a project lasts a few weeks, it is
possible to remember all the steps without carefully recording and explaining
them.  If you have a small dataset and a handful of tiny functions, it is
reasonable to throw them all in a directory and post them online.  As datasets
get larger and the analysis more complex, it quickly becomes impossible to
keep track of what you have done without carefully organizing and
recording your work.  Large datasets and complex analyses are typical in research.

Second, we intended to teach the students efficient reproducible practice with
the standard tools that experts use for this purpose
\cite{millman2014developing}. Our own practice has taught us that the power
of these tools only becomes clear when you use them to do substantial work in
collaboration with your peers.  In contrast, we have found that teaching
``easy'' tools that need less initial investment has the paradoxical effect of
making it harder for students to move on to the more powerful and efficient
tools that they will need for their daily work.

\section{Material and Methods}\label{methods}

We taught \emph{Reproducible and Collaborative Statistical Data
Science},\footnote{\url{http://www.jarrodmillman.com/stat159-fall2015/}} during
the fall semester of 2015.  The course was offered through the department of
Statistics at UC Berkeley, and was open to upper-level undergraduates (as STAT
159) as well as graduate students (as STAT 259).

The course entry requirements were Berkeley courses STAT 133 (Concepts in
Computing with Data), STAT 134 (Concepts of Probability), and STAT 135
(Concepts of Statistics).  Together these courses provide basic
undergraduate-level familiarity with probability, statistics, and statistical
computing using the R language.
Many students were from statistics and/or computer science; other majors
included cognitive science, psychology, and architecture.
During the 15-week long semester, we had three hours of class time per week in
two 90-minute sessions, plus two hours of lab time.  Students were expected to
work at least eight hours per week outside class.  Project reports were due in
week~17, two weeks after the last class.

\subsection{Course overview}

This was the entry for our course in the Berkeley course catalog:

\begin{quote}
A project-based introduction to statistical data analysis. Through case
studies, computer laboratories, and a term project, students learn
practical techniques and tools for producing statistically sound and
appropriate, reproducible, and verifiable computational answers to
scientific questions. Course emphasizes version control, testing,
process automation, code review, and collaborative programming.
Software tools include Bash, Git, Python, and \LaTeX.
\end{quote}

\subsubsection{Tools and process}

We had three guiding principles in our choice of tools and process:
(a) to teach efficient reproducible practice with standard expert tools,
(b) to teach fundamental mathematical and statistical principles using
explanation with simple code, and
(c) students should be able to build their own analysis from basic building
blocks.

Applying these principles, we taught command line tools, version control, and
document machinery using text files.  Rather than focusing on a specific
neuroimaging analysis package, we taught scientific coding with Python and its
associated scientific libraries \cite{millman2011python, perez2011python}, and
then showed the students how to run standard statistical procedures on imaging
data using these tools \cite{millman2007analysis}.

\blockpar{Command line.}
The Unix environment is the computational equivalent of the scientists'
workbench \cite{preeyanon2014reproducible}.  The Unix command line, and the
Bash shell in particular, provides mature, well-documented tools for building
and executing sequences of commands that are readable, repeatable, and can be
stored in text files as scripts for later inspection and execution.  Quoting
\cite{wilson2014best}---the Bash shell makes it easier to ``make the computer
repeat tasks'' and ``save recent commands in a file for re-use.''

The graphical user interface of operating systems such as Windows and macOS
can obscure the tree structure of the file system, making it harder to think
about the organization of data as a hierarchy of directories and files.  The
command line tools make the file system hierarchy explicit, and so make it
easier to reason about data organization.

\blockpar{Version control.}
Version control is a fundamental tool for organizing and storing code and
other products of data analysis.
Distributed version control allows many people to work in parallel on the
same set of files.
The tools associated with distributed version control make it easier for
collaborators to review each other's work and suggest changes.

Git is the distributed version control system that has become standard in
open source scientific computing. It is widely used in industry, with automated
installers for all major operating systems.
Web platforms such as GitHub,\footnote{\url{https://github.com}}
Bitbucket,\footnote{\url{https://bitbucket.org}} and
GitLab\footnote{\url{https://about.gitlab.com}} provide web interfaces to
version control that simplify standard methods of collaboration such as code
review, automated testing, and issue tracking (see below).

\blockpar{Scientific Python.}
Python is a general purpose programming language,
popular for teaching and prevalent in industry and academia.
In science, it has particular strength in
astronomy, computational biology, and data science.
Its impact in scientific computing rests on a stack of popular scientific
libraries including NumPy (computing with arrays), SciPy (scientific
algorithms including optimization and image processing), Matplotlib (2D
plotting), Scikit-Learn (machine learning), and Pandas (data science).  The
Nibabel library\footnote{\url{http://nipy.org/nibabel}\label{nibabel}} can
load files in standard brain image formats as arrays for manipulation and
display by the other packages.

\blockpar{Peer review.}
Regular peer review is one of the most important ways of learning
to be an effective author of correct code and valid data analysis.  Git and its
various web platforms provide a powerful tool for peer review.

\emph{Pull requests} are a version control interface feature pioneered by
GitHub.  A pull request presents proposed changes to the shared
repository in a convenient web interface. Collaborators can comment on the
changes, ask questions, and suggest improvements.  The discussion on this
interface forms a record of decisions, and a justification for each accepted
set of changes.

GitHub, like other hosting platforms for version control, provides an
interface for creating \emph{issues}.  These can be reports of errors
that need to be tracked, larger scale suggestions for
changes, or items in a to-do list.

\blockpar{Functions with tests.}
Functions are reusable blocks of code used to perform a specific task.
They usually manipulate some input argument(s) and return some output
result(s).
There are several advantages to organizing code into functions rather
than writing stream-of-consciousness scripts.
Functions encapsulate the details of a discrete element of
work.  A top-level script may call a well-named function to express a
particular step, allowing the user to ``hide'' the mechanism of this step in
the function code.  A top-level script that uses functions in this way gives a
better overview of the structure of the analysis.
Functions make it easier to reuse code at various points in the
analysis, instead of rewriting or copying and editing it.
Finally, writing functions allows the programmer to test small parts of the
program in isolation from the rest.

Testing is an essential discipline to give some assurance to the authors and
other users that the code works as expected.
Tests should also document how the code is expected to be used.
Inexperienced coders typically greatly underestimate how often they make
errors; expecting, finding, and fixing errors is one of the foundations of
learning from continued practice.  There are tools that can measure what
proportion of the lines of code are covered by the tests---this is \emph{test
coverage}.

\blockpar{Continuous integration.}
Tests are useless if they are not run, or if the authors do not see the
results.  Modern code quality control includes \emph{continuous integration
testing}, a practice that guarantees that the tests are run frequently and
automatically, and the results are reported back to the coding team.
Fortunately, this practice has become so ubiquitous in open source, that there
are large free services that implement continuous integration, such as Travis
CI,\footnote{\url{https://travis-ci.org}\label{travis-ci}} Circle
CI,\footnote{\url{https://circleci.com}} and
Appveyor.\footnote{\url{https://www.appveyor.com}}  These services can
integrate with hosting sites like GitHub in order to run tests after each
change to the code.

\blockpar{Markup languages.}
A markup language allows you to write content in plain text, while also
maintaining a set of special syntax to annotate the content with structural
information.
In this way, the plain text or \emph{source} files are human-readable and work
well with standard version control systems.
To produce the final document, the source files must undergo a \emph{build} or
\emph{render}
step where the markup syntax (and any associated style files) are passed to a
tool that knows how to interpret and render the content.

\LaTeX\,and Markdown represent two extremes of markup languages, each
with their own usefulness.
\LaTeX\,is notation-heavy, but powerful.
In contrast, Markdown is notation-light, but limited.

Pandoc is a command line document processor that can convert between multiple
markup formats.  It can also generate rendered output from source files with
text markup.  For example, it can generate PDF files from text files written
with Markdown markup.

\blockpar{Reproducible workflows.}
When working with the Unix command line, we frequently generate files
by performing a sequence of commands.
The venerable Make system was originally written to automate the process
of compiling and linking programs, but is now widely used to
automate all types of command line workflows.

Makefiles are machine-readable text files consisting of rules
specifying the sequence of commands necessary for generating
certain files and for tracking dependencies between files.
Consider the following Makefile rule:
\begin{verbatim}
    progress.pdf: progress.md
            pandoc -t beamer -s progress.md -o progress.pdf
\end{verbatim}
This rule defines the procedure to generate the PDF slide show file
\texttt{progress.pdf}. The first line specifies the \textit{target} of the
rule, and any \textit{prerequisites}.  Here the target is
\texttt{progress.pdf}; building this target requires the source
Markdown text file \texttt{progress.md}. The indented line gives the
\textit{recipe}, which is the mechanism by which the target
should be generated from the prerequisites.  In this case, the recipe is to
execute the \texttt{pandoc} utility on the Markdown file, with various options
applied.\footnote{\url{http://pandoc.org/MANUAL.html\#producing-slide-shows-with-pandoc}}
If you edit \texttt{progress.md}, you can regenerate \texttt{progress.pdf}
from the command line using \texttt{make progress.pdf}.

In the example above, the recipe was a single command, but Makefile recipes
often involve several commands.  The target of a rule is often a filename
(e.g., \texttt{progress.pdf}), but can also be an arbitrary name to describe
the sequence of commands in the recipe.  The dependencies of a rule can be
filenames (e.g., \texttt{progress.md}), but may also be other rules in the
Makefile.  These features allow Makefiles to chain together complex sequences
of commands necessary for generating multiple target files, which may depend on
other files or steps; such chaining is important for building and maintaining
reproducible workflows.

\subsubsection{Lectures, labs, notes, homeworks, readings, and quizzes}

The weekly lectures and labs prepared students with the skills and
background needed for the group project.
We demonstrated methods and tools during lecture and lab,
and expected students to do additional research and reading as they worked on
the group project.
For example, we only gave one or two lectures on each of Unix, Git, Python,
testing, and NumPy.
Each weekly lab focused on the software stack from previous lectures.
For example, after the Git lecture, we used Git and GitHub to fetch and submit
all exercises;
after introducing code testing in lecture, we continually reinforced testing in
the labs.
Course work outside labs also made heavy use of the toolstack, including each
new tool as it was introduced in lectures.

For the first three lectures, we introduced students to the Unix environment
and version control using Git.
The fourth lecture was a high-level introduction to neuroimaging and
functional magnetic resonance imaging (or FMRI).
We then introduced students to scientific computing with Python.
We referred students to our course notes on
Bash,\footnote{\url{http://www.jarrodmillman.com/rcsds/standard/bash.html}}
Git,\footnote{\url{http://www.jarrodmillman.com/rcsds/standard/git-intro.html}}
Make,\footnote{\url{https://www.youtube.com/watch?v=-Cp3jBBHQBE}}
Python,\footnote{\url{http://www.jarrodmillman.com/rcsds/standard/python.html}}
scientific Python packages,\footnote{\url{http://www.scipy-lectures.org/}}
and \LaTeX.\footnote{\url{https://www.youtube.com/watch?v=8khoelwmMwo}}

By the ninth lecture (week~5), we started to focus on the analysis of FMRI data
with Python.
Data analysis lectures continued to develop the students' skills with
Unix and the scientific Python software stack through practical problems,
which arose as we taught FMRI analysis methods.

We assigned two homeworks.
Students had two weeks to work on each homework.
Homeworks were assigned and submitted using Git and private GitHub repositories during
the first half of the semester.
Typically, assignments were given as a collection of tests that tested the
desired behavior of named functions, and function templates with missing
implementations.
The functions were documented according to the NumPy documentation
standard.\footnote{\url{https://github.com/numpy/numpy/blob/master/doc/HOWTO_DOCUMENT.rst.txt}}
Students added implementations to the functions according to the documentation
and ran the tests to check that their code returned the correct results
for the provided tests.
Assignments were graded using extended tests, which were not provided to
the students before they submitted their work.
The homework reinforced the material we taught during the
beginning of the course and focused on scientific programming
in Python.

Over the course of the semester, seven readings were assigned on a roughly
bi-weekly basis.
The readings consisted of articles that emphasized the core concepts
of the class, either with respect to scientific computing, or neuroimaging.
Students were required to compose a two-paragraph write-up that both summarized
the main points of the article, and commented on it.

The labs and quizzes emphasized hands-on experience with the computing tools
and process associated with reproducible and collaborative computing (e.g.,
version control, \LaTeX, etc.).
Multiple choice quizzes were held at the beginning of lab sessions, and
emphasized the computing aspects of the course material as
opposed to the statistical and neuroimaging components covered in the lectures.
The remainder of the lab was devoted to providing hands-on experience via
collaborative work on breakout exercises.
These exercises were formulated as small projects to be worked on
in groups of three to four students.

We also used the labs to help the students practice critical code review.
For example, we had the students form small groups and review one another's
solutions to the second homework.
This involved all the students in each group creating pull requests to a
shared GitHub repository with their solutions.
Then they had to use the GitHub code review model, which we had been practicing
in class, to review three functions in each solution.
We asked them to summarize their findings with respect to
clarity, brevity, and performance.
We used their summaries to inform a subsequent discussion about code review.

\subsection{Course project}\label{project}

The course centered around a semester-long group project that accounted for 55\%
of the overall course grade.
Students worked on their projects in teams for three months.

Early in the semester (week~5) groups of three to five students formed teams.
To make sure every aspect of the project work was computationally
reproducible and collaborative, we immediately created a publicly visible
project repository for each team.
The groups defined the scope of their projects iteratively; the teams
submitted their proposals in week~6, a third of the way through the semester.
We gave each team feedback to clarify project motivation and goals.

\begin{figure}
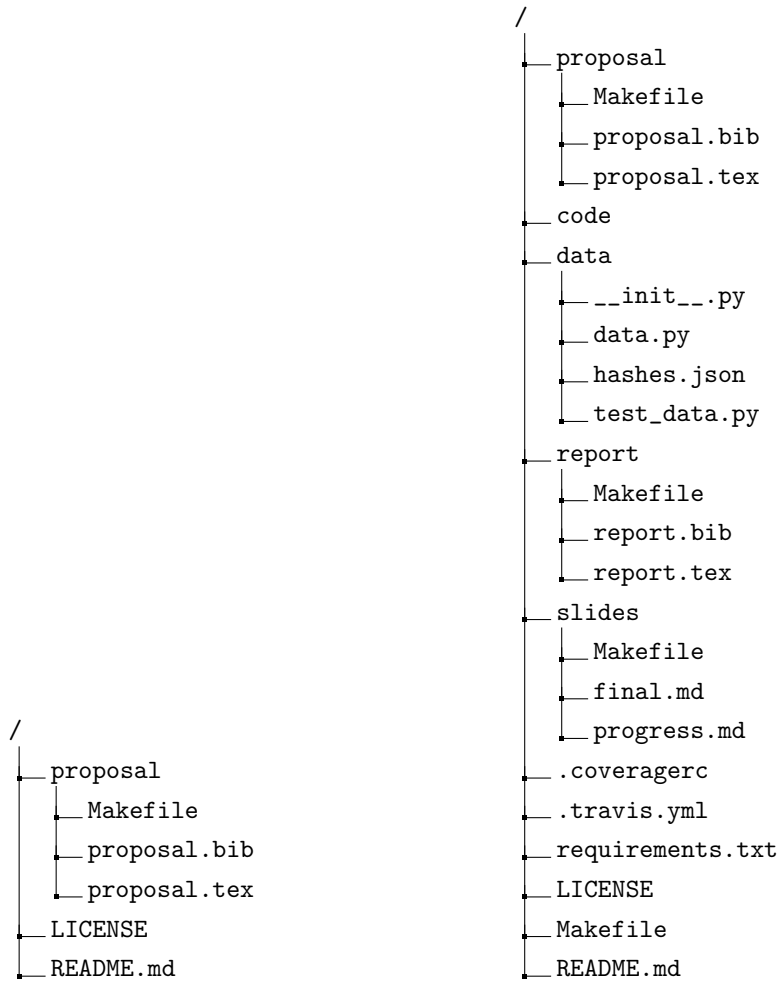

\centering
\begin{minipage}[b]{0.4\textwidth}
\dirtree{%
.1 /.
.2 proposal.
.3 Makefile.
.3 proposal.bib.
.3 proposal.tex.
.2 LICENSE.
.2 README.md.
}
\subcaption{Initial repository template.}\label{fig:init-repo}
\end{minipage}
\begin{minipage}[b]{0.4\textwidth}
\dirtree{%
.1 /.
.2 proposal.
.3 Makefile.
.3 proposal.bib.
.3 proposal.tex.
.2 code.
.2 data.
.3 \_\_init\_\_.py.
.3 data.py.
.3 hashes.json.
.3 test\_data.py.
.2 report.
.3 Makefile.
.3 report.bib.
.3 report.tex.
.2 slides.
.3 Makefile.
.3 final.md.
.3 progress.md.
.2 .coveragerc.
.2 .travis.yml.
.2 requirements.txt.
.2 LICENSE.
.2 Makefile.
.2 README.md.
}
\subcaption{Final project template.}\label{fig:final-repo}
\end{minipage}
 \caption{The initial and final repository directory template for student
    projects.  We gave the students a project copied from the initial template
    in week~5, from which they would write and build and their initial
    proposal.  As the course progressed, we pushed two updates to their
    repository. The first gave them a template for building their final
    report, their top-level \texttt{README.md} file and file specifying an
    open license for their work. The second gave them: an initial set of
    functions we had shown them in class with associated tests (in the
    \texttt{code} directory, not shown); machinery to trigger automatic
    execution of tests on Travis-CI servers; automated code coverage
    reporting; and a template for writing slides for their first progress
    report.}
\label{fig:repo}
\end{figure}

For their project proposals, we required the teams to use a \LaTeX\,template,
which we provided, and to add the source file(s) to their project repository
(see Figure~\ref{fig:repo}).
Each proposal involved
(a) identifying a published FMRI paper and the accompanying data,
(b) explaining the basic idea of the paper in a paragraph,
(c) confirming that they could download the data and that what they
downloaded passed basic sanity checks (e.g., correct number of subjects), and
(d) explaining what approach they intended to take for exploring
the data and paper.
All teams chose a paper and accompanying dataset from
OpenFMRI,\footnote{\url{https://www.openfmri.org/}} a publicly-available
depository of MRI and EEG datasets
\cite{poldrack2013toward,poldrack2015openfmri}.

We reviewed the project proposals and met with each team in week~7 to help
them refine their ideas.
The feedback process continued throughout the semester, with students
regularly submitting drafts according to milestones defined within the project
timeline.
As much as possible, we used this process to replicate the way we
work with collaborators in the lab.

Project feedback included peer-review, where each research team was
responsible for cloning another team's project, running the tests and
analyses, and evaluating the resulting technical document.
The peer-review process proved particularly valuable, as the students 
benefited greatly from the exposure to the coding techniques and 
organizational principles of their peers.

Students used GitHub's pull request mechanism to review all project work.
Code was written as a collection of functions (with tests) and short
scripts that called these functions to perform the project data analysis.
We taught students to test their code thoroughly.
We set up each project with a configuration file for the Travis
CI continuous integration service, and enabled integration of
this service with their GitHub repository.  The configuration file specified
that their tests would be run and reported on the Travis CI servers each
time they changed their code on the GitHub hosting site.
We added automated code coverage testing; each change to the code on GitHub
resulted in a measure of the proportion of all code lines covered by tests.
We expected students to keep measured test coverage above 90\%.
While each team decided when to merge pull requests, we recommended that
all pull requests should have at least three participants, test coverage
should not decrease, and tests should pass before being merged.

Students also used the GitHub issues interface to record project discussions
and actions taken by the group members.
We told students to create a pull request
or issue with code and text and use the GitHub interface to ask one or more
of us to respond, if they had any questions about their projects.
Periodically, we would review their work and open issues.

We covered this workflow in depth during lectures and in lab,
and made clear that the project grade would be based on
(a) the final report,
(b) the analysis code,
(c) whether we could run their analysis code to reproduce all
their results and figures, and
(d) how effectively the team collaborated using the techniques
taught in the course (e.g., pull requests, code review, testing).
We told the students that we would use the pull request and issue
discussions as evidence for their contributions to the project, and as data for
their final grades.

Rather than have students commit the raw FMRI data to their project repository, we had
them commit code to download and validate
the raw data used in their project (week~8).
Since this was the first code they added to their repositories, we gave them example
code to download a file and validate it by checking the file against a stored
checksum of the file content.
Once they added code to handle downloading and validating the data,
we had them commit code for their exploratory data analysis (week~9).
After this, the only milestones we provided were for progress
presentations and report drafts.

\begin{figure}
\centering
\begin{verbatim}
    Progress reports and presentations
    ----------------------------------
    
    1. Briefly describe data
       - Paper
       - dsnum from openfmri.org
       - What kind of data is it?
    
    2. Briefly describe what you've done so far
       - data fetching/preprocessing
       - initial analysis
       - plots? figures?
    
    3. What is your plan?
       - What analysis will you perform?
       - How does it relate to the original data analysis?
         - Will you use all the data?  Why or why not?
         - What model/preprocessing steps will you simplify?
       - What are the problems you face?
         - Try to be explicit about your issues?
         - Suggest potential solutions and/or approaches.
         - What do you need to research more?  Have you found sources?
         - Will you try to make "inferences" from the data?
           - How will you deal with multiple comparisons?
           - How will you attempt to validate your model?     
    
    4. Process
       - What was the hardest part of the process so far?
         - Git workflow, Python, fMRI data, all of the above
         - Having an ill-defined assignment?
       - How successful have you been at overcoming these obstacles?
       - What issues have you faced working as a team?  How have you
         been addressing them?
       - What parts of the class have been the most useful?
       - What parts have been the least helpful or most confusing?
       - What do you need to do to successfully complete the project?
       - How difficult are you finding it to make your work reproducible?
         Do you feel confident that you are in a position to make your
         projects reproducible?
       - What would be most helpful for your team in the remaining weeks?
         Additional lectures?  Structured or unstructured group work?
       - What potential topics would be most useful for us to cover?
         - Overview of brain / neuroanatomy?
         - More linear regression (ANOVA)? PCA? The mathematics or the implementation?
         - Machine learning (classification, prediction, cross-validation)?
         - Permutation tests (and maybe bootstrap)?
         - Software tools (Git, Make, Python, statmodels, etc.)
         - Technical writing and scientific visualization?
         - Advanced topics (regularized regression, selective inference)?
\end{verbatim}
 \caption{Class lecture material giving instructions for the first progress
    report.  This is the text we showed them during class, and discussed, to
    prepare them for their first progress report.  As for almost all our
    lecture material, we posted this text to the class website.}
\label{fig:progress}
\end{figure}

The students gave their first progress report and presentation in week~12;
it consisted of a description of the data they were using, what they had done
so far, their plan for completing the project, and a reflection on the overall
process (see Figure~\ref{fig:progress}).
We required the groups to write their slides in Markdown text format for
building into PDF with Pandoc.  They committed the Markdown source to the
project repository.
The students gave their second slide presentation in week~15, two weeks before
the projects were due, discussing their current progress and their plan for
finishing their project.
At the end of the project, in week~17, they had to commit the final PDF of
their reports and provide instructions for generating the report PDF
from the source \LaTeX\,files.

Early in the project, we told the students that we would grade their project
work on whether it was reproducible.  In order to reproduce their work, we
told the students we would blindly follow the instructions in a text file
named \texttt{README.md} in the root directory of the project repository.
When we graded their projects, the \texttt{README.md} file had to explain how
to install, configure, and run all aspects of their project.
Each \texttt{README.md} had to specify how to rebuild all the components of
the project and in what order to do so.
For example, it might include a section specifying
the \texttt{make} commands to execute, such as:
\begin{verbatim}
    make data      # download the data
    make validate  # ensure the data is not corrupted
    make eda       # generate figures from exploratory analysis
    make analysis  # generate figures and results
    make report    # build final report
\end{verbatim}

\subsection{Neuroimaging data analysis}\label{analysis}

In terms of neuroimaging, our aims were for students to
(a) understand the basic concepts in neuroimaging,
and how they relate to the wider world of statistics, engineering, and computer science;
(b) be comfortable working with neuroimaging data and code, so they could write
their own basic algorithms, and understand other people's code;
(c) continue to use the computational techniques we had taught them to improve
efficiency and help their understanding.

For this course we concentrated on the statistical analysis of FMRI data using
a linear model.  We designed each lecture to teach the next analysis step the
students would need for their project work.

All the following teaching used simple Python code to show them how the
mathematics works in practice, and to give them sample code that they could
edit and run for themselves.\footnote{For an example lesson showing our use
of Python code to illustrate the mathematical and statistical ideas,
see \url{http://www.jarrodmillman.com/rcsds/lectures/glm_intro.html}.}
We specifically avoided using off-the-shelf imaging analysis software packages,
and encouraged the students to build their own analyses from the building
blocks we had given them.  In lectures, we interleaved teaching with short
group exercises where students took the code from class and extended it to
solve a new problem.

We covered the following topics:

\begin{itemize}

\item
    The idea of images as visual displays of values in arrays.
\item
    The standard neuroimaging NIfTI image format as a simple image container,
        combining metadata (such as shape) and image pixel / voxel data stored
        as a sequence of numbers.  We used the Nibabel Python package
        to load NIfTI data as three-dimensional arrays.
\item
    Four dimensional (4D) images as sequential time-series of 3D image
        volumes.  Slicing arrays to get individual volumes from the 4D
        time-series.  Extracting single-voxel time-courses from 4D images.
\item
    Building regressors for the statistical model.  We introduced the idea of
        the neuronal time-course model as the hypothesized change in neuronal
        activity caused by the task.  For a block design, this model becomes a
        box-car, with all rest periods having identical low activity, and all
        activity periods having identical high activity.  The FMRI acquisition
        measures something like blood flow, rather than neuronal activity, but
        blood flow changes slowly in response to changes in neuronal activity.
        We introduced the linear-time-invariant assumption in predicting the
        hemodynamic activity from the hypothesized neuronal activity, and then
        the hemodynamic response function as the invariant response.  Finally
        we demonstrated the mechanism of convolution as a numerical method to
        express the effect of the hemodynamic response function acting in a
        linear-time-invariant way, and the caveats to these assumptions.  We
        now had hemodynamic regressors to fit to our voxel time-course data.
\item
    Correlation as a simple test of linear association between the hemodynamic
        regressor and voxel signal.  Calculating correlation with a single
        regressor for every brain voxel and the idea of a statistical
        parametric map.
\item
    The linear model as a matrix formulation of multiple regression.  We started with
        simple (single-regressor) regression as another measure of linear
        association.  We expressed simple regression as a linear model and
        showed how the model can be expressed as the addition of vectors and
        vector / scalar multiplication.  This leads to the matrix formulation
        of simple regression, and thence to multiple regression.  We introduce
        dummy indicator variables to express group membership and show how
        these relate to group means.  We showed with code how this mathematics
        can express statistical methods that they already know, such as
        regression, $t$-tests, and ANOVA.
\item
    High-resolution sampling of regressors.  The simple cases that we had
        covered above assumed events or blocks that started at the same time
        as the scanner started to collect a new brain volume.  This is not the
        case in general.  To deal with events that do not start at volume
        onsets, we need to generate a hemodynamic time-course at higher
        time resolution than the spacing of the scan onsets, and then sample
        this regressor (using linear interpolation) at the scan onset times.
\item
    Parametric modulation of event regressors.  Some of the OpenFMRI datasets
        that the students had chosen used regressors to model parametric
        modulation of events.  There is one regressor to model the hemodynamic
        effect of a particular event type on average, and another to capture
        the variation of the hemodynamic event activity as a linear function
        of some characteristic of that event, such as duration or intensity.
\item
    Spatial smoothing with a Gaussian kernel; smoothing as a form of
    convolution; the \texttt{scipy.ndimage} subpackage in SciPy as an
    implementation of Gaussian and other smoothing.

\item
    The idea of voxel and millimeter coordinates in brain images, and the
    image affine as a mapping between them.  The students needed this in order
    to relate the coordinates reported in their papers to voxels in their own
    analyses.

\end{itemize}

We reinforced the material from class lectures and exercises in the homeworks.
Figure~\ref{fig:diagnosis_script} shows an excerpt from the second homework,
in which we asked to students to fill out functions implementing various
algorithms on 4D FMRI data and use these functions to build up a simple
diagnostic test for outlier volumes in the time series.  Finally, they
applied a linear (multiple regression) model to the data to show that removing
outlier scans reduced residual variance from the model fit.

\begin{figure}
\centering
\begin{Verbatim}[commandchars=\\\{\}]
\PY{l+s+sd}{\PYZdq{}\PYZdq{}\PYZdq{} Script to run diagnostic analysis on FMRI run}

\PY{l+s+sd}{The FMRI \PYZsq{}run\PYZsq{} is a continuous collection of one or more 3D volumes.}
\PY{l+s+sd}{A run is usually stored as a 4D NIfTI image.}
\PY{l+s+sd}{Fill in the code necessary under the comments below.}
\PY{l+s+sd}{\PYZdq{}\PYZdq{}\PYZdq{}}

\PY{l+s+sd}{\PYZdq{}\PYZdq{}\PYZdq{}}
\PY{l+s+sd}{* Load \PYZdq{}ds114\PYZus{}sub009.nii\PYZdq{} as an image object}
\PY{l+s+sd}{* Load the image data from the image}
\PY{l+s+sd}{* Drop the first four volumes, as we know these are outliers}
\PY{l+s+sd}{\PYZdq{}\PYZdq{}\PYZdq{}}

\PY{l+s+sd}{\PYZdq{}\PYZdq{}\PYZdq{}}
\PY{l+s+sd}{Use your vol\PYZus{}std function to get the volume standard deviation values}
\PY{l+s+sd}{for the remaining 169 volumes. Save these values to a text file}
\PY{l+s+sd}{called \PYZsq{}vol\PYZus{}std\PYZus{}values.txt\PYZsq{}.}
\PY{l+s+sd}{\PYZdq{}\PYZdq{}\PYZdq{}}

\PY{l+s+sd}{\PYZdq{}\PYZdq{}\PYZdq{}}
\PY{l+s+sd}{Use the iqr\PYZus{}outlier detection routine to get indices of outlier volumes.}
\PY{l+s+sd}{Save these indices to a text file called \PYZsq{}vol\PYZus{}std\PYZus{}outliers.txt\PYZsq{}.}
\PY{l+s+sd}{\PYZdq{}\PYZdq{}\PYZdq{}}

\PY{l+s+sd}{\PYZdq{}\PYZdq{}\PYZdq{}}
\PY{l+s+sd}{Plot all these on the same plot:}
\PY{l+s+sd}{* The volume standard deviation values;}
\PY{l+s+sd}{* The outlier points from the std values marked with an \PYZsq{}o\PYZsq{} marker;}
\PY{l+s+sd}{* A horizontal dashed line at the lower IRQ threshold;}
\PY{l+s+sd}{* A horizontal dashed line at the higher IRQ threshold;}

\PY{l+s+sd}{Save the figure to the current directory as ``vol\PYZus{}std.png``.}
\PY{l+s+sd}{\PYZdq{}\PYZdq{}\PYZdq{}}

\PY{l+s+sd}{\PYZdq{}\PYZdq{}\PYZdq{}}
\PY{l+s+sd}{Next calculate and plot the RMS difference values:}
\PY{l+s+sd}{* Calculate the RMS difference values for the image data;}
\PY{l+s+sd}{* Use the ``iqr\PYZus{}outlier`` function to return indices of possible}
\PY{l+s+sd}{  outliers in this RMS difference vector;}

\PY{l+s+sd}{On the same plot, plot the following:}
\PY{l+s+sd}{* The RMS vector;}
\PY{l+s+sd}{* The identified outlier points marked with an `o` marker;}
\PY{l+s+sd}{* A horizontal dashed line at the lower IRQ threshold;}
\PY{l+s+sd}{* A horizontal dashed line at the higher IRQ threshold;}

\PY{l+s+sd}{Save this plot as ``vol\PYZus{}rms\PYZus{}outliers.png``}
\PY{l+s+sd}{\PYZdq{}\PYZdq{}\PYZdq{}}
\end{Verbatim}
 \caption{Above we list the first few lines of \texttt{diagnosis\_script.py}.
This file was part of the second homework, which focused on detecting outlier
3D volumes in a 4D FMRI image.  Tasks included:
(a) implementing functions on image arrays using NumPy,
(b) exploring FMRI data for outliers,
(c) running least-squares fits on different models, and
(d) making and saving plots with Matplotlib.}\label{fig:diagnosis_script}
\end{figure}

We encouraged students to use the basic Python building blocks for their
project analyses, but we did not insist.  Our rule was that, if they used
other software, they had to persuade us that they had a sound understanding of
the algorithms that the other software was using.

\subsection{Project grading}

We based the evaluation of the final projects on criteria that emphasized the
underlying principles of reproducibility, collaboration, and technical
quality.
See Table~\ref{tab:rubric} for the final project grading rubric, which we
gave to the students on week~11---the week before the first progress
presentations.
From the perspective of reproducibility, we evaluated projects on whether the
presented results could be generated from their repositories according to the
documentation they provided in their \texttt{README.md} text file.  When we
could not reproduce the analysis, we raised one or more GitHub issues to
negotiate with the project team.
We graded the code tests with respect to code coverage.
Grading for the collaborative aspects of the project used the information
from the project history provided by the Git version history and GitHub web
artifacts, including code contributions, as well as reviewing GitHub
pull requests and discussion on GitHub issues.
Finally, we assessed the technical quality of the project in terms of
the clarity of the final report and with respect to how well the proposed
goals of the study were met by the final results of the analysis.

\begin{table}
\centering
\begin{tiny}
\begin{tabular}{|>{\bf}c|>{\raggedright}p{4cm}|>{\raggedright}p{4cm}|>{\raggedright\arraybackslash}p{4cm}|}
\hline
 & 
\multicolumn{1}{c|}{\textbf{\ding{51}-}}
 & 
\multicolumn{1}{c|}{\textbf{\ding{51}}}
 & 
\multicolumn{1}{c|}{\textbf{\ding{51}+}}
 \\
\hline

Questions
 & 
Questions overly simplistic, unrelated, or unmotivated
 & 
Questions appropriate, coherent, and motivated
 & 
Questions well motivated, interesting, insightful, and novel
 \\
\hline

Analysis
 & 
Choice of analysis overly simplistic or incomplete
 & 
Analysis appropriate
 & 
Analysis appropriate, complete, advanced, and informative
 \\
\hline

Results
 & 
Conclusions missing, incorrect, or not based on analysis

Inappropriate choice of plots; poorly labeled plots; plots missing
 & 
Conclusions relevant, but partially correct or partially complete

Plots convey information but lack context for interpretation
 & 
Relevant conclusions tied to analysis and context

Plots convey information correctly with adequate and appropriate reference information
 \\
\hline

Collaboration
 & 
Few members contributed substantial effort or each members worked on only part of project
 & 
All members contributed substantial effort and everyone contributed to all aspects of project
 & 
All members contributed substantial effort to each project aspect
 \\
\hline

Tests
 & 
Tests incomplete, incorrect, or missing
 & 
Tests cover most of the project code
 & 
Extensive and comprehensive testing
 \\
\hline

Code review
 & 
Pull requests not adequately used, reviewed, or improved
 & 
Pull requests adequately used, reviewed, and improved
 & 
Code review substantial and extensive
 \\
\hline

Documentation
 & 
Poorly documented
 & 
Adequately documented
 & 
Well documented
 \\
\hline

Readability
 & 
Code readability inconsistent or poor
 & 
Code readability consistent and good quality
 & 
Code readability excellent
 \\
\hline

Organization
 & 
Poorly organized and structured repository
 & 
Reasonably organized and clear structure
 & 
Elegant and transparent code organization
 \\
\hline

Presentation
 & 
Verbal presentation illogical, incorrect, or incoherent

Visual presentation cluttered, disjoint, or illegible

Verbal and visual presentation unrelated
 & 
Verbal presentation partially correct but incomplete or unconvincing

Visual presentation is readable and clear

Verbal and visual presentation related
 & 
Verbal presentation correct, complete, and convincing

Visual presentation appealing, informative, and crisp

Verbal and visual presentation clearly related
 \\
\hline

Writing
 & 
Explanation illogical, incorrect, or incoherent
 & 
Explanation correct, complete, and convincing
 & 
Explanation correct, complete, convincing, and elegant
 \\
\hline

Reproduciblity
 & 
Code didn't run
 & 
Makefile recipes fetch data, validates fetched data, generates all results and figures in report
 & 
Makefiles generate EDA work and supplementary analysis
 \\
\hline

\end{tabular}
 \caption{Project grading rubric.
An ``A'' was roughly two or more check pluses and no check minuses.}
\label{tab:rubric}
\end{tiny}
\end{table}

\section{Results}\label{results}

There were a total of eleven research teams composed of three to five
students, each responsible for completing a final project of their own design
using datasets available through the OpenFMRI organization.  We named groups
arbitrarily with Greek letters (e.g., alpha, kappa, zeta).  The project
repositories are public at \url{https://github.com/berkeley-stat159}.

Although we allowed students to select any OpenFMRI dataset, in fact all
groups selected one of the following:

\begin{itemize}

\item
    \texttt{ds000005}:\footnote{\url{https://openfmri.org/dataset/ds000005}}
        \textit{Mixed-gambles task} \cite{tom2007neural} (groups delta,
        epsilon, eta, theta);

\item
    \texttt{ds000009}:\footnote{\url{https://openfmri.org/dataset/ds000009}}
        \textit{The generality of self-control} \cite{cohen2014generality}
        (group alpha);

\item
    \texttt{ds000105}:\footnote{\url{https://openfmri.org/dataset/ds000105}}
        \textit{Visual object recognition} \cite{haxby2001distributed}
        (groups kappa, zeta);

\item
    \texttt{ds000113}:\footnote{\url{https://openfmri.org/dataset/ds000113}}
        \textit{A high-resolution 7-Tesla FMRI dataset from complex natural
        stimulation with an audio movie} \cite{hanke2014high} (groups beta,
        lambda);

\item
    \texttt{ds000115}:\footnote{\url{https://openfmri.org/dataset/ds000115}}
        \textit{Working memory in healthy and schizophrenic individuals}
        \cite{repovs2012working} (groups gamma, iota).

\end{itemize}

During grading, we succeeded in fully replicating all group analyses.  The
minor problems that arose were largely platform differences between macOS used
by the students and the Linux system we were using for grading.  Four projects
reproduced without issue, six required us to raise one GitHub issue, and one
required two issues.

We counted the lines of Python code using the \texttt{cloc}
utility.\footnote{\url{http://cloc.sourceforge.net}}  The median lines of code
across projects was 1,494 (range 624--8,582), of which tests comprised
(median; range) 440; 36--592.  As this suggests, we were not successful in
persuading the students to test a large proportion of their code.  We
estimated coverage of all lines of Python code at a median of 12.4\% (range
2.0--91.8).\footnote{See
\url{https://github.com/jarrodmillman/rcsds-paper/tree/master/project_metrics}
for details of this estimate.}  The code coverage machinery we had put in
place did not measure coverage of code outside expected code directories, and
the students had put a large proportion of their code outside these
directories, leaving the automated code coverage score above 90\% in all
cases, even though a large proportion of the code was not in fact covered by
tests.

No team used SPM, FSL, or AFNI.  Two projects imported the
\texttt{nilearn} package,\footnote{\url{https://nilearn.github.io}} which is a
Python neuroimaging package designed for machine learning, but both projects
used Nilearn only for spatial smoothing and basic volume visualization.  One
project used a single function from the
\texttt{Dipy} package\footnote{Dipy is a package for analysis of diffusion
imaging data: \url{http://nipy.org/dipy}} for calculating a
within-brain mask for the functional images.  Another used a class from the
\texttt{NiTime} package\footnote{NiTime is a library for time-series analysis
of data from neuroscience experiments: \url{http://nipy.org/nitime}} for
temporal filtering.  All projects used the basic NumPy, SciPy, and Matplotlib
packages; eight of eleven used the \texttt{Scikit-Learn}
package\footnote{\url{http://scikit-learn.org}} for machine learning; six used
the \texttt{Statsmodels} package\footnote{\url{http://www.statsmodels.org}}
for statistical modeling.  The uses of Statsmodels were: AutoRegressive
Integrated Moving Average (ARIMA) modeling of voxel time-series and
autocorrelation plots (two projects); logistic regression (two projects);
ordinary least squares modeling (one project); White test for
Heteroscedasticity (one project); mixed effects modeling (one project).  Of
these, we had only shown the implementation of ordinary least squares analysis
in class.

There was a wide range in the novelty and scope of the projects.  Most
projects consisted of a serious attempt to replicate one or several
statistical findings in the original paper, with the addition of some further
extension or exploration.  These extensions were typically the application of
other statistical techniques.  For example, project epsilon was one of the
four groups working on the dataset ds000005---the \textit{Mixed-gambles
task}.  They first explored the imaging data using the outlier detection
machinery that all students had developed in the second homework.  Next they
explored various logistic regression models of the behavioral data.  For the
imaging data, they used the SciPy \texttt{ndimage} subpackage to smooth the
data, as we had briefly shown in class, and then followed some hints in the
lectures to explore different confound models such as linear and quadratic
drift, and Principal Components as regressors.  They used code
from class to implement the general linear model at each voxel, and calculate
$t$- and $p$-values using contrast vectors.  Finally they thresholded their images
using Bonferroni correction.

Other projects did more substantial technical or intellectual extensions of
the original analysis.  Project alpha explored an analysis from
ds000009---\textit{The generality of self-control}.  Although we had not covered this in
class, they discovered from data exploration and reflection that the times of
slice acquisition would affect their statistical modeling, and developed code
to shift their model back and forth in time corresponding to the time of slice
acquisition.  They tried various confound models on the data, including linear
drift, Fourier bases and different numbers of PCA regressors, and explored
these models with selection criteria such as the Bayes and Akaike Information
Criteria, and adjusted $R^2$.  They ran diagnostics to detect voxels violating
assumptions of normality. For whole brain analysis, they implemented the FDR
multiple comparison correction and tried other methods for identifying
activated brain regions, including hierarchical clustering.  Finally, they
experimented with ARIMA models of the voxel time course.

Project lambda did an heroic effort to replicate the analysis of ds000113---a
high-resolution FMRI dataset of subjects listening to a description of the
film Forrest Gump.  This dataset had many technical challenges; the images are
unusually large, at over 1 million voxels per volume, and of unusually long
duration, at around 450 volumes per run, and having 8 runs per subject.  The
analysis requires the correlation of many voxel time-courses in each subject
with voxel time-courses in all other subjects in the analysis.  The group did
a variety of explorations, including finding some artifacts in the original
data, and an error in the published paper, and then went on to replicate the
original correlation analysis, on a smaller number of subjects.  They ran the
analysis on Amazon Web Services machines in order to deal with the demands of
processing time and memory.  As they note in their \texttt{README.md} ``We
strongly encourage running on a machine with 120 GBs of accessible RAM to
emulate development environment.'' They extended the analysis by running a
random forest model to detect volumes corresponding to outdoor and indoor
scenes in the film.  With all this, they achieved around 92\% code test
coverage.

Seventeen out of forty undergraduates and six out of ten graduate students
completed anonymous course evaluations.  The primary question of interest to
us was ``Considering both the limitations and possibilities of the subject
matter and the course, how would you rate the overall effectiveness of this
course?''.  Ratings were on a 1 through 7 scale with 4 corresponding to
``moderately effective'' and 7 corresponding to ``extremely effective.''
The average undergraduate and graduate scores were
4.71 and 6.0 respectively, against a department average across all courses of
5.23.  Undergraduate, graduate, and department average ratings for ``I enjoyed
this class'' on a 1 through 7 scale (with 4 corresponding to
``somewhat'' and 7 to ``very'')
were 4.69, 6.00, and 4.96 respectively.

70.5\% of the undergraduate respondents and all the graduate respondents
claimed to have worked 10 or more hours a week on average.

\section{Discussion}\label{discussion}

Most neuroimaging researchers agree that computational reproducibility is
desirable, but rare.  How should we adapt our teaching to make reproducibility
more common?

The usual practical answer to this question, is that we should train
neuroimaging and other computational researchers as we do now, but add short
courses or bootcamps that teach a subset of the tools we taught here but
without the substantial practice.  That is, reproducibility is an addition on
top of current training.

We believe this approach is doomed to failure, because the students are
building on an insecure foundation.  Once you have learned to work in an
informal and undisciplined way, it is difficult to switch to a process that
initially demands much more effort and thought.  Rigor and clarity are hard to
retrofit.
To quote the American chemist Frank Westheimer:
``A couple of months in the laboratory can frequently save a couple of hours
in the library.''

For these reasons, our course took the opposite approach.
We put a substantial, collaborative, and open-ended project at the center of
the course.
We then started with the tools and process for working with numerical data, and
for building their own analyses.
We used this framework as a foundation on which to build their understanding of
the underlying ideas.
As we taught these tools, we integrated them into their exercises and homework,
and made it clear how they related to their project work.

Our claim is that this made our teaching and our students much more efficient.
The secure foundation made it easier for them to work with us and with each
other. As they started their project work, early in the course, they could
already see the value of these tools for clarity and collaboration. Our
students graduated from the course with significant experience of using the
same tools that experts use for sound and reproducible research.

\subsection{Did we really teach neuroimaging analysis?}

By design, our course covered tools and process as well as neuroimaging.  We
used neuroimaging as the example scientific application of these tools.  Can
we claim to have taught a meaningful amount of neuroimaging in this class?

Class content specific to neuroimaging was a guest lecture in class 4 (of 25),
and teaching on the brain images, correlation, and the general linear model
from classes 9 through 15.  We covered only the standard
statistical techniques needed for a basic analysis of a single run of FMRI
data.  This is a much narrower range than a standard neuroimaging course, but
we covered these topics in much greater depth than is typical for a
neuroimaging course.  This was the basis for final projects that were
substantial and well-grounded.

Typical imaging courses do not attempt to
teach the fundamental ideas of linear models, but assume this understanding
and move onto imaging specifics.  This assumption is almost entirely false for
neuroscience and psychology students, and largely false for our own
students, even though most had training from an undergraduate
statistics major. As a result of this incorrect assumption, it is rare for
students of neuroimaging to be confident in their understanding of the
statistics they are using.  We taught the linear model from the first
principles of vector algebra, using the tools they had just learned, to build
a simple analysis from basic components.  As a result, when the students got
to their projects, they had the tools they needed to build their own
neuroimaging analysis code, demonstrating and advancing their own
understanding.

We note the difficulty of the task that we gave the students, and the
extent of their success.  We made clear that their project was an open-ended
exploration of an FMRI dataset and paper.  Few students had any
experience or knowledge of FMRI before the course. The only guidance we gave
was that they should prefer well-curated datasets from the OpenFMRI
depository.  In order to design and implement their project, they had to
understand at least one published FMRI paper in some depth, with limited
assistance from their instructors.  We gave no example projects for them to
review, or templates for them to follow.  The submitted projects were all serious efforts to reproduce and
/ or extend the published results, and all included analysis code that they
had written themselves.

We did not teach the full range of neuroimaging analysis.  For example, we did
not cover pre-processing of the data, random effects analysis, or inference
with multiple comparisons.   Our claim is that what we did teach was a sound
and sufficient foundation from which they could write code to implement their
own analyses.  This is a level of competence that few neuroimagers achieve
after considerable experience.

\subsection{Did we really teach computational reproducibility?}

We stressed the importance of reproducibility throughout the course, and
made it an explicit feature for grading of the final project, so we were not
surprised to find that it was possible to reproduce all of the final projects
with little extra clarification or fixes from the group members.

By design, we gave the students a project that was close to a real
scientific data analysis.  They had to work with a messy and real dataset to
explore the data, define their problem and solve it with an analysis that they
implemented themselves, to various degrees.  We required them to work closely
in teams, often remotely, using standard tools for collaboration.

We found this combination of a substantial analysis problem, expert tools, and
the requirement for reproducibility, was effective at giving the students a
concrete sense of the difficulties in making an analysis reproducible, and how
these can be overcome.  We speculate, from our own work, that the
experience of building a reproducible analysis makes it easier to commit to
the tools and practice needed for reproducible work in the future.

\subsection{Did we teach the right tools?}

We do not believe that the individual tools we chose were controversial. We
taught the tools that we use ourselves, and that we would teach to
students working with us.

We could have used R instead of Python, but Python is a better language for
teaching, and has better libraries for neuroimaging and machine learning.

A more interesting question is whether we went too far in forcing the students
to use expert tools.  For example, we required them to write their
report in \LaTeX, and their presentation slides and analysis description in
plain text with Markdown markup.  We asked them to do all project interaction
using GitHub tools such as the issue tracker and pull requests.  We set and
marked code exercises with Git version control and text files, rather than
interactive alternatives, such as Jupyter Notebooks.

Of course---at first---some students complained that they would rather use
Facebook for code discussions and PowerPoint for their presentation slides.
Should we have allowed this?  We believe not.  Our own experience
of using these tools is that their power only becomes apparent
when you learn to use them for all the tasks of analysis, including generating
reports and presentations.  Mixing ``easy'' but heavy tools like PowerPoint
and Facebook with ``simple'' and light tools like text editors and Markdown
causes us to lose concentration as we switch modes from heavy to light and
back again. It is easy to be put off by the difficulty of getting used to the
light tools, and therefore fail to notice the clarity of thought and
transparency of process that they quickly bring.  Successfully switching from
heavy to light is a process that requires patience and support; it is best
done in the context of a class where there are examples of use from coursework
and support from experienced instructors who use these tools in their daily
work.

\subsection{Do students need to be trained in programming?}

One common justification for not training students in good coding practice is
the assertion that scientists do not need to be programmers.  However, scientists do
have to use, read, and write code, so a more defensible
statement would be that scientists do not need to be \emph{good} programmers.
It is surely true that successful scientists can be bad programmers, but bad
programmers are inefficient and prone to error; they are less likely to detect
errors, and improve very slowly over time.  We should invest teaching time
to help our students work efficiently and continue learning for the rest of
their careers.

\subsection{Do students need a substantial, collaborative, and open-ended project?}

We put great emphasis on the final project in this course, and this was clear
to most of the students.  In response to the evaluation survey question
\emph{``What advice would you give to another student who is considering taking this course?''}
one undergraduate wrote:
\begin{quotation}
``[U]nlike most group projects (which last for maybe a few weeks tops or
could conceivably be pulled off by one very dedicated person), this one will
dominate the entire semester. . . . Try to stay organized for the project and
create lots of little goals and checkpoints. You should always be working on
something for the project, whether that's coding, reviewing, writing, etc. Ask
lots of questions and ask them early!''
\end{quotation}

The size of the project meant that the students had to learn to
collaborate with each other efficiently, often remotely, as students had
different class schedules.  Many of the tools that we taught, such as
distributed version control, only become essential when working in
collaboration.  Conversely, if you are not collaborating with others, it can
be difficult to see why it is worth investing the time to understand powerful
tools like the Git version control software, or the GitHub interface.

Working in collaboration, and working reproducibly, changes the way that we
think.  If we have to explain our work to others in the group, or to
another user of the work, then we develop the expectation that whatever we
write will always be something we will demonstrate and explain to others.  It
becomes part of the work to communicate our ideas and explain what we did.

It was important that the project was open-ended.  If the student has to solve
a small problem with a single correct answer, they can often check
whether they have the right answer, and do not need to worry about the quality
of the process that generated it.  In an open-ended project, it is likely that
the group will need to explore different analyses as they progress.  The
answers are not known, and the group has to proceed with care, to avoid making
false conclusions.  This is typical of real scientific analysis, and puts a
higher burden on rigor and testing, than a typical small classroom problem.

We should emphasize how hard it was to get the students to engage with the
project early, and work steadily.  In the first class, we gave the students a
document describing the project, including the various project deadlines.  We
continued to emphasize the project and project deadlines in announcements.
Nevertheless, it was only half way through the course that the students began
to realize how open-ended their task was, and how much work they would have to
do.  For a few weeks, the class was anxious, and our job as instructors was to
project faith in the teams' ability to define a tractable and interesting
question.  We mention this to say that, a large open-ended project has many
advantages, but to make it work, it does take courage from the students
and the instructors.

\subsection{What background do students need?}

The requirements for our course were previous classes on probability,
statistics, and the use of the R programming language.

We would be happy to relax the requirement for probability.  We did not refer
to the ideas of probability in any depth during the course.  Authors JBP and
MB have taught similar material to neuroscience and psychology students who
lack training in probability; the pace of teaching and level of understanding
were similar for the statistics students in this course and the other
students we have taught.

Psychology and neuroscience students do have some statistical training.  Our
impression was that this background was necessary for us to be able to as
start as quickly as we did in the analysis of linear regression.

We would also keep the requirement for some programming experience. We assumed
familiarity with programming ideas such as for loops, conditionals, and
functions.  In our psychology / neuroscience courses that used similar
material, we required some experience of programming in a language such as
Python, R, or MATLAB.  It is possible that a brief introduction would be
enough to fulfill this requirement, such as a bootcamp.

\subsection{Could this material be covered by a bootcamp or hack week?}

There are several existing programs designed to address the problems
of informal and inefficient computational practice.  The best known may be
Software Carpentry \cite{wilson2014software}.  A Software Carpentry course
consists of a two-day bootcamp covering many of the same tools we used,
including the Unix command line, distributed version control, and programming
with a high-level language such as Python or R. Hack weeks can be a related
approach to the same problems. Hack weeks vary in format from a series of
tutorials to a week of collaborative work on a project that has already been
agreed \cite{huppenkothen_hack_2017}.  The tutorials and modeled working
practice usually include the same set of tools we used.

These approaches have some similarity with the early classes in our course,
but there are differences.  Bootcamps and hack weeks attract volunteers with
interest in or commitment to efficient and reproducible practice. These are
typically graduate students or post-docs that already have some technical
experience.  They are motivated enough to come to campus on a weekend,
stay longer at a conference, or travel some distance.  Even so, we suspect
that it would be possible to show that short introductions in
bootcamps will not be effective in changing practice in the medium term,
without later reinforcement and support by peers.

Our students did choose our course from others they could have taken, but most
of them had little background of good practice in computation.  The majority
were undergraduates.  We suspect that many of the students finishing our
course did have enough experience of using the tools they had used, to
continue using them in their daily work, and teach others to do the same.  We
base this suspicion on the depth and quality of the project work.

Bootcamps and hack weeks can be useful, but their starting point is an
attempt to augment an aging curriculum that does not recognize the need for
training in accurate, effective, and reproducible computation.  We should fix
this by changing our curriculum; learning to work this way takes time,
practice, and support, and we teach it best with substantial commitment from
students and instructors.

\subsection{Where would such a course fit in a curriculum?}

Our course would not fully qualify a student for independent research in
neuroimaging.  As we discuss above, we did not cover important aspects of
imaging, including spatial and temporal pre-processing of data, random effects,
or control of false positives in the presence of multiple comparisons.  Where
should the elements of our course sit in relation to a larger program for
training imagers and other computational scientists?

We think of this course as a foundation.  Students graduating from this course
will be more effective in learning and analysis.  The tools that they have
learned allow the instructor to build analyses from basic blocks, to
show how algorithms work, and make them simple to extend.  We
suggest that a course like ours should be first in a sequence, where the
following courses would continue to use these tools for exposition and for
project work.

A full course on brain imaging might start with an introduction to Python and
data analysis, possibly as a week-long bootcamp at the start of the semester.
Something like our course would follow.  There should be follow-up courses
using the same tools for more advanced topics such as machine learning,
spatial pre-processing, and analysis of other modalities. We suggest that each
of these courses should have group project work as a large component, in which
the students continue to practice techniques for efficient reproducible
research.

\subsection{What factors would influence the success of such a course?}

We should note that there were factors in the relative success of this
course that may not apply in other institutions.

Berkeley has as a strong tradition in statistical computing and
reproducibility.  Leo Breiman was a professor in the Berkeley statistics
department, and an early advocate of what we would now call data science.  He
put a heavy emphasis on computational teaching to statisticians.  Sandrine
Dudoit is a professor in the statistics and biostatistics departments, and a
founding core developer of the Bioconductor project devoted to reproducible
genomics research in R.  The then head of the statistics department, Philip
Stark, is one of many Berkeley authors (including KJM) to contribute to the
recent book ``The Practice of Reproducible Research''
\cite{kitzes2017practice}. Authors KJM, MB, and JBP have worked with Mark
D'Esposito, who runs the Berkeley Brain Imaging Center, and was an early
advocate for data sharing in FMRI.  Other imaging labs on campus take this
issue seriously and transmit this to their students.  If
there had not been such local interest in the problem of reproducibility,
students may have been less convinced of the importance of
working in a reproducible way, especially given the short-term convenience
of a less disciplined working process.

Students of statistics and other disciplines at Berkeley are well aware of the
importance of Python in scientific computing, and in industry.  Tech firms
recruit aggressively on campus, and Python is a valuable skill in industry.
The cross-discipline introductory course in Fundamentals of Data Science uses
Python. The new Berkeley Institute of Data Science has a strong emphasis
on Python for scientific computing.

In the same way, a booming tech sector nearby made it more obvious to students
that they would need to learn tools like Git and GitHub.
For example, public activity on GitHub is one
feature that companies use to identify candidates for well-paid positions as
software engineers.

We required students to use \LaTeX\,to write their final reports.  This was an
easy sell to statistics students, but students outside mathematical fields
might be less agreeable; if we were teaching psychology and neuroscience
students, we might choose another plain text markup language, such as
Markdown, with the text files run through the Pandoc document processor to
write publication quality output.

\section*{Conclusion}

Our course differs from other imaging courses that we know of, in several
respects.  We started with correct computational practice, before teaching
neuroimaging.  We emphasized computation as a way of explaining the underlying
ideas and taught the fundamentals of the linear model, rather than assuming
students had this background, or would get this background later.
We made these ideas concrete with a substantial open-ended project designed to
be as close as possible to the experience of graduate research.

We were largely successful in teaching the students the tools they can continue
to use productively for collaborative and reproducible research.

We are sure that most of our readers would agree that, in an ideal world, we
should teach students to work in this way, but, given all the other
classes our students must take, can we justify the time and energy that this
course needs?  We believe so, but we know that not all our readers will be
convinced.  This is true of students as well as
instructors.  We put such emphasis on the final project, precisely because we
know that it is difficult to see how important these tools are, if you
have not used them, or used them only in toy projects.  We cannot easily teach
these tools to our fellow teachers, so instead we offer this sketch of an
experiment, to make the discussion more concrete.  Imagine we took 200
students, and randomized them into two groups of 100 each.  The first group
takes something like the course we describe here as an introduction, and then
one or more further courses on imaging.  The second does not take such a
course, but instead has more traditional teaching on imaging that covers a
wider range of techniques, using off the shelf imaging software.  Part of
their later teaching would include some techniques for reproducible research.
Two years after such an experiment, we predict that
students from the first group will have a greater understanding of what they
are doing, will be more effective in analysis, and more likely to
experiment with new ideas.  We think it much more likely that the first group
will be doing reproducible research.

\section*{Conflict of Interest Statement}

The authors declare that the research was conducted in the absence of any
commercial or financial relationships that could be construed as a potential
conflict of interest.

\section*{Author Contributions}

KJM was the lead instructor and was responsible for the syllabus and project timeline;
KJM and MB were responsible for lectures and created homework assignments;
KJM and RB were responsible for labs, readings, quizzes, and grading;
all authors held weekly office hours to assist students with their projects.
KJM and MB wrote the first draft of the manuscript;
all authors wrote sections of the manuscript, contributed to manuscript revision, 
as well as read and approved the submitted version.

\section*{Funding}

KJM was funded in part by the Gordon and Betty Moore Foundation through Grant
GBMF3834 and by the Alfred P. Sloan Foundation through Grant 2013-10-27 to the
University of California, Berkeley.

\section*{Acknowledgments}
We would like to thank the students who took our class for their enthusiasm,
hard work, and thoughtful feedback.  Russ Poldrack and the OpenFMRI team
were the foundation that made it possible for us to design the student
projects as we did. Russ and the team were very generous in providing
processed data at short notice for the students to analyze.  Alexander Huth
gave an inspiring overview of neuroimaging analysis in a guest lecture.
St{\'e}fan van der Walt, Fernando P{\'e}rez, and Paul Ivanov have through long
collaboration and numerous discussions helped clarify and refine our approach
to teaching.  Thanks to Mark D'Esposito for his steadfast support and
encouragement as we experimented with new ways of teaching.
Finally, we would like to thank Mark D'Esposito, Sandrine Dudoit, and
Nima Hejazi for their constructive feedback on a draft of this paper.

\bibliographystyle{plain}

\begin{thebibliography}{WAB{\etalchar{+}}14}

\bibitem{buckheit1995wavelab}
Jonathan~B Buckheit and David~L Donoho.
\newblock Wavelab and reproducible research.
\newblock In {\em Wavelets and statistics}, pages 55--81. Springer, 1995.

\bibitem{cohen2014generality}
Jessica~R Cohen and Russell~A Poldrack.
\newblock Materials and methods for openfmri ds009: The generality of self
  control.
\newblock
  \url{https://openfmri.org/media/ds000009/ds009_methods_0_CchSZHn.pdf}, 2014.
\newblock Accessed March 3, 2018.

\bibitem{huppenkothen_hack_2017}
Daniela Huppenkothen, Anthony Arendt, David~W. Hogg, Karthik Ram, Jake
  VanderPlas, and Ariel Rokem.
\newblock Hack {Weeks} as a model for {Data} {Science} {Education} and
  {Collaboration}.
\newblock {\em arXiv:1711.00028 [astro-ph, physics:physics]}, October 2017.
\newblock arXiv: 1711.00028.

\bibitem{hanke2014high}
Michael Hanke, Florian~J Baumgartner, Pierre Ibe, Falko~R Kaule, Stefan
  Pollmann, Oliver Speck, Wolf Zinke, and J{\"o}rg Stadler.
\newblock A high-resolution 7-tesla fmri dataset from complex natural
  stimulation with an audio movie.
\newblock {\em Scientific data}, 1:140003, 2014.

\bibitem{haxby2001distributed}
James~V Haxby, M~Ida Gobbini, Maura~L Furey, Alumit Ishai, Jennifer~L Schouten,
  and Pietro Pietrini.
\newblock Distributed and overlapping representations of faces and objects in
  ventral temporal cortex.
\newblock {\em Science}, 293(5539):2425--2430, 2001.

\bibitem{kitzes2017practice}
Justin Kitzes, Daniel Turek, and Fatma Deniz, editors.
\newblock {\em The practice of reproducible research: case studies and lessons
  from the data-intensive sciences}.
\newblock University of California Press, 2017.

\bibitem{millman2011python}
K~Jarrod Millman and Michael Aivazis.
\newblock Python for scientists and engineers.
\newblock {\em Computing in Science \& Engineering}, 13(2):9--12, 2011.

\bibitem{millman2007analysis}
K~Jarrod Millman and Matthew Brett.
\newblock Analysis of functional {Magnetic Resonance Imaging in Python}.
\newblock {\em Computing in Science \& Engineering}, 9(3):52--55, 2007.

\bibitem{millman2014developing}
K~Jarrod Millman and Fernando P{\'e}rez.
\newblock Developing open-source scientific practice.
\newblock {\em Implementing Reproducible Research. CRC Press, Boca Raton, FL},
  pages 149--183, 2014.

\bibitem{poldrack2013toward}
Russell~A Poldrack, Deanna~M Barch, Jason~P Mitchell, Tor~D Wager, Anthony~D
  Wagner, Joseph~T Devlin, Chad Cumba, Oluwasanmi Koyejo, and Michael~P Milham.
\newblock {Toward open sharing of task-based fMRI data: the OpenfMRI project}.
\newblock {\em Frontiers in Neuroinformatics}, 7, 2013.

\bibitem{poldrack2015openfmri}
Russell~A Poldrack and Krzysztof~J Gorgolewski.
\newblock {OpenfMRI: open sharing of task fMRI data}.
\newblock {\em NeuroImage}, 2015.

\bibitem{perez2011python}
Fernando P{\'e}rez, Brian~E Granger, and John~D Hunter.
\newblock Python: an ecosystem for scientific computing.
\newblock {\em Computing in Science \& Engineering}, 13(2):13--21, 2011.

\bibitem{preeyanon2014reproducible}
Likit Preeyanon, Alexis~Black Pyrkosz, and C~Titus Brown.
\newblock Reproducible bioinformatics research for biologists.
\newblock {\em Implementing Reproducible Research}, pages 185--216, 2014.

\bibitem{repovs2012working}
Grega Repovs and Deanna~M Barch.
\newblock Working memory related brain network connectivity in individuals with
  schizophrenia and their siblings.
\newblock {\em Frontiers in human neuroscience}, 6:137, 2012.

\bibitem{tom2007neural}
Sabrina~M Tom et~al.
\newblock The neural basis of loss aversion in decision-making under risk.
\newblock {\em Science}, 315(5811):515--518, 2007.

\bibitem{wilson2014best}
Greg Wilson, DA~Aruliah, C~Titus Brown, Neil P~Chue Hong, Matt Davis, Richard~T
  Guy, Steven~HD Haddock, Kathryn~D Huff, Ian~M Mitchell, Mark~D Plumbley,
  et~al.
\newblock Best practices for scientific computing.
\newblock {\em PLoS Biol}, 12(1):e1001745, 2014.

\bibitem{wilson2014software}
Greg Wilson.
\newblock Software carpentry: lessons learned.
\newblock {\em F1000Research}, 3, 2014.

\end{thebibliography}

\newcommand{\etalchar}[1]{$^{#1}$}

\end{document}